\documentclass[12pt,a4paper]{article}

\usepackage{amsmath,amssymb,amsthm,graphicx,color,bm,soul}
\usepackage{hyperref}

\begin{document}

\title{Recollections about V\"axj\"o conferences. Preface to the special issue  ``Quantum Information and Probability: from Foundations to Engineering'' (QIP23)} 
\author{Andrei Khrennikov\\ 
Linnaeus University, International Center for Mathematical Modeling\\  in Physics and Cognitive Sciences
 V\"axj\"o, SE-351 95, Sweden}

\date{}                     

\maketitle

\abstract{As the preface to the special issue for the conference ``Quantum Information and Probability: from Foundations to Engineering'' (QIP23), I wrote these notes with recollection about V\"axj\"o conferences. These conferences covered 25 years of my life (2000-24) and played the crucial role in evolution of my own views on the basic problems of quantum foundations. I hope to continue this conference series as long as possible. Up to my understanding, this is the longest conference series on quantum foundation in the history of quantum physics. These notes contain my recollections of conversations with the world's leading experts on quantum foundations. I think that such notes may have the historical value. My own views on quantum foundations are specific and they evolved 
essentially during  25 years. Finally, I discovered the practically forgotten pathway in physical foundations developed by von Helmholtz, Hertz, Boltzmann, and Schr\"odinger and known as the Bild conception. A scientific theory is a combination of two models, observational and causal. Coupling between these models can be tricky. A causal model can operate with hidden quantities which can't be identified with observables. From this viewpoint, Bell's coupling of subquantum and quantum models is very special and the violation of the Bell inequalities doesn't close (local) ways beyond quantum mechanics.}

{\bf keywords:} quantum foundations; V\"axj\"o conferences; quantum information revolution;  quantum-like revolution; foundational debates 

\section{V\"axj\"o series of annual conferences on quantum foundations}

The quantum information revolution, also known as the second quantum revolution,  has influenced  not only the development of quantum technology, but it also has been accelerating studies on foundations of quantum theory. Regardless the future perspectives of wide implementation, quantum computers, networks, and cryptography are new powerful systems for testing quantum foundations and elaborating
novel insights on theoretical description of micro and macro-phenomena and their interrelation. One of the main consequences of the quantum information revolution is that quantum systems are treated as information processors nowadays. Roughly speaking, the main theoretical interest is to quantum soft-ware. Of course, the experimental implementation is based on advanced hardware, including highly efficient photo-detectors. 

The Nobel Prize (2022) given to Aspect, Clauser, and Zeilinger ``for experiments with entangled photons, establishing the violation of Bell inequalities and pioneering quantum information science'' was generated by the quantum foundational studies. And I am proud that the series of the conferences on quantum foundations, information, and probability  arranged annually in V\"axj\"o (Sweden) since year 2000 contributed essentially to critical analysis of the``Bell project'' \cite{Bell2,Bell1}. The hot debates on the possible interpretations of quantum mechanics (QM) and especially on the violation of the Bell type inequalities were supplemented by the talks of leading experimentalists involved in testing of these inequalities (e.g., Aspect, de Martini, Genovese, Gisin, Grangier, Hanson, Ramelow, Shalm, Zeilinger, Weihs, Weinfurter,...). Such annual assembles energized studies on this topic, both theoretical and experimental. One of my organizational contributions to this project was arranging the special session {\it Big event – Final Bell test} at the conference Quantum and Beyond in 2016. The session included the talks of the representatives of all experimental groups performed the loophole free experiments. The session was concluded by the talk of Zeilinger who presented the future of experiments in quantum foundations.

\begin{itemize}
\item  A. Aspect (the talk was given by P. Grangier): Closing the door on Einstein and Bohr's quantum debate.
\item G. Weihs: Violation of Bell’s inequality under strict Einstein locality conditions.
\item R. Hanson, From the first loophole-free Bell test to a quantum Internet.
\item  M. Giustina and M. Versteegh: Significant-loophole-free test of local realism with entangled photons.
\item  K. Shalm: A strong loophole - free test of Bell's inequalities. 
\item  H. Weinfurter: Event ready loophole free Bell test using heralded atom-atom entanglement.
\item P. Grangier: Violation of Bell's Inequalities in a quantum realistic framework.
\item  A. Zeilinger: The future of Bell experiments. 
\end{itemize}

Videos of these talks can be found at lnuplay, see \cite{VAspect,VWeihs,VHanson,VGiustina,VWeinfurter,VZeilinger}. 
It is important to note that not all participants of the conference agreed with Aspect and Grangier  that  the Einstein and Bohr debate is closed. During the following conferences the debates on the possible interpretations of the outputs of these loophole free experiments were continued. This is the good place to recall that Bell by himself predicted this situation \cite{Bell1}.

By discussing the completion of the project on the loophole free Bell tests we should recall the contributions  of Gill, Glancy, and Larsson who made crucial contributions into statistical justification of these tests and defended their statistical legacy during
confrontation with the ``anti-Bell opposition'' which was widely represented in V\"axj\"o. Personally I think that the existence of such opposition played the crucial role in the evolution of the Bell project and igniting general interest to Bell inequalities.
The most active  representatives of the ``anti-Bell opposition'' in V\"axj\"o were  Accardi, Adenier, De Raedt and  Michielsen, Hess and Philipp, Khrennikov, Kupczynski, Nieuwenhuizen. The views of Nobel Prize Laureate ´t Hooft also deviated from the conventional interpretation of the violation of the Bell inequality, but in V\"axj\"o he didn't present them openly (see below my recollection of our conversations).  

By completing this part of discussion the Bell project and the the Nobel Prize 2022, I would like to highlight once again 
the contribution of Gill, Glancy, and Larsson: their mathematical (probabilistic) analysis  was crucial in completion of this project.
Unfortunately, the role of mathematics in the experiments is typically dimmed.  
\\

I have personally been very engaged in the foundational research and I enjoyed the V\"axj\"o conferences, the talks and discussions with talented people.\footnote{Who are generally talented not only in science. I think that many participants remember the great evening when Daniel Greenberger played piano in Teleborg Castle.} Only one big dark cloud has always been at the quantum horizon, the cloud of quantum nonlocality. Spooky action at a distance disturbs, but at the same time stimulates my mind.
\\

This metaphor ``dark cloud
at the quantum horizon'' I borrowed from a lunch-talk with Anton Zeilinger, where he told the story about how he became interested in foundations. In 1950s he participated in a conference which assembled the world-leading experts in quantum physics. The general attitude of participants was that development of quantum theory is more or less completed. There was just one dark cloud at the quantum horizon, the cloud of quantum gravity. As I understood, he was curious whether there are other clouds at the quantum foundational horizon and this curiosity led him to the Nobel Prize in 2022. 

 In this nonlocal atmosphere I tried to seek salvation in the circle of QBists led by Fuchs with active support of Mermin and Schack 
(see \cite{Fuchs1}-\cite{Fuchs7} for some references on QBism, especially Fuchs' papers for V\"axj\"o proceedings).
Although QBism was widely represented at the conferences in V\"axj\"o since year 2002 and I have always supported plenary talks and special sessions of QBists, I am a strong antagonist  of QBism for the subjective probability interpretation of QM and highlighting the personal experiences of ``quantum agents''. Educated within the Soviet probability school and being lucky to have conversations with Kolmogorov on foundations of probability, I treat the probability objectively. The first time I heard  about subjective probability from Fuchs and hearing such nonsense annoys me a lot. However, QBism declares that in QM there is no spooky action at a distance, QM is  local. So, may be even QBism is better than nonlocality! 

This was precisely H\"ansch's message in his V\"axj\"o talk. I remind that he received the Nobel Prize (2005) for  ``contributions to the development of laser-based precision spectroscopy, including the optical frequency comb technique''. Hence, he is excellent experimentalist and his appeal to QBism is not due to attraction by the  subjective viewpoint on QM. He told that he needs a consistent local interpretation of QM and QBism provides it. H\"ansch's talk induced hot debates and he was attacked by ``quantum realists''. I remember the critical comments of Vaidman and especially Elitzur. At one of the conferences, we had the social opinion poll on acceptance of QBism as the basic interpretation of QM. Surprisingly many world's leading experimentalists in quantum foundations (e.g.,  Weihs) supported QBism. QBism was less popular among theoreticians and philosophers, it was rejected by the majority of the participants. I suspect that Zeilinger is sympathetic to QBism, although he did not declare this openly, but  his talk in Vienna at the conference ``Quantum [Un]Speakables II: 50 Years of Bell's Theorem'' was so much in the spirit of QBism.

I spent a lot of time by trying to proceed in quantum foundations and handle the violation of the Bell inequalities without appealing to nonlocality. I have understood that one of the barriers on the way  towards liberation of QM from nonlocality is operating with the term ``local realism''. This is one of the most misleading notions in the history of physics, but it is firmed in the heads of the majority of physicists who are interested in fundamental questions. Two totally different notions, namely, realism and locality, are jointly mixed and one of them shadows another.

The first signal of this notion ambiguity I received from the answer of Gisin to my comment to his talk that the notion of quantum nonlocality is misleading.  Gisin answered that the most misleading is the notion of local realism and first one should eliminate it from discussions on violation of the Bell inequalities: one can not keep one term and dismiss the other. Local non-realism doesn’t mean anything, as Gisin wrote in article \cite{Gisin} (see also my note \cite{Ka}). I remark that Bell did not operate with the notion of local realism, at least in his book \cite{Bell1} summarizing his main works he did not even mention it.  

In principle, one can reject both realism and locality. In particular, one famous experimentalist who is well known for the rejection of realism (as many representatives of the Vienna's circle)  often told to me that QM would finally stimulate people to reject realism, but once he suddenly added that today it is clear that one should  also reject locality. The later statement was really sudden for me in the light of all our previous conversations. I ignore the double rejection possibility simply by using Occam's razor. But I remember Cromwell's rule and I may be mistaken. By himself Gisin is definitely in favor of rejection of locality.

Similar conversation was with famous supporter  of the many worlds interpretation. 
He always positioned himself as realist (being in one of numerous worlds). But once he told that QM is nonlocal. I was really shocked, I didn't expect from him such hit below the belt. When I became a bit relaxed, I asked him: ``Why?!'' And he pointed to nonlocality of the projection postulate. 

During twenty three V\"axj\"o conferences I have debated quantum foundations with many bright scientists and often we disagreed, but this was in the spirit of these conferences. I am especially thankful to the members of the ``V\"axj\"o club'', those who participated in the majority of the conferences, especially to Bengtsson, D'Ariano, Fuchs, Jaeger, Elitzur, Larsson, Nieuwenhuizen, Plotnitsky, Rauch, Vaidman, and Weihs.

At the beginning of this century Bohmians, especially Goldstein and Hiley, strongly influenced my realist attitude - my  first conference had the title ``Bohmian Machanics - 2000''. I was impressed by spiritual strength of Basil Hiley who often struggled alone against  a hundred of anti-Bohmians. This was a good lesson of scientific struggle. I and other participants enjoyed the hot battle between Basil  Hiley and Marlan Scully on whether Bohmian mechanics can be experimentally distinguished from conventional QM. Only strong pressure from the chairman, Theo  Nieuwenhuizen, led to termination of this battle.

As well the Bohmian interpretation, the many worlds interpretation was not so widely represented in V\"axjö. But during 20 years 
Lev Vaidman advertized it  enthusiastically. This interpretation didn't  meet so strong criticism as the  Bohmian one, may be because 
the majority of participants didn't take it seriously. When I heard the speech of Lev at one of the first conferences (year 2002?), 
I was really shocked and looked at Chris Fuchs who adviced to invite Lev. But Chris nicked - ``it is ok!'' Later 
I started to think that the many worlds interpretation is not so bad, as one of the consistent alternatives to quantum nonlocality.    
The main problem for me is still understanding of which experiment outcomes lead to world's split? And the personal contacts with Lev
only increased my disappointment. During long time, I was sure that each random experiment splits the world (where it is performed).
Then once I, Lev, and my daughter Polina (who was teenager  at that time) walked near the lake from the university  to the downtown. 
Polina  was curious in the many worlds story that Lev told with enthusiasm. Suddenly she said :``Lev, if I unexpectedly push you into this lake, then you will only fall there in one of the worlds?'' Lev looked at her with caution (you can expect anything from a teenager) and (unexpectedly for me) said, that such a splitting of worlds is impossible for macro-objects of such mass as he is; for example, with an amoeba it would be possible...     
\\ 

Coming to V\"axj\"o of Gerard `t Hooft was the breath of the fresh air for realists. His braveness
in struggle against the orthodox Copenhagen views, multiplied by the weight of the Nobel Prize, induced the wave of enthusiasm.
It was impression that his journey beyond quantum would soon lead to establishing of the realist interpretation of quantum theory.

Unfortunately - or fortunately? - it didn't happen. The first signal that ´t Hooft  proceeds too quickly without taking care of
many basic foundational problems came during one of our conversations, it happens that he ``didn't pay attention'' that his model confronts with the Bell theorem. But his reaction to these news was plane and it seems that this theorem didn't disturb him so much.
(I was really surprised!) In the post-conference email exchange his attitude was Bohm-like and, as I understood, nonlocality was not a substantial problem. Later he came with the argument based on superdeterminism and, in particular, rejection of free will. From my present perspective, in principle `t Hooft need not appeal to  superdeterminism. The latter, although logically possible, is not so attractive for majority of the physicists. The model of `t Hooft matches perfectly the Bild conception of Herz-Botzmann-Schr\"odinger on the interrelation between observational and causal theoretic models (see section \ref{2}). Coupling between two levels of the physical description can have a variety of colors. As was already repeated a few times, on the way beyond quantum one need not follow Bell with his specific rules for such coupling.

Contacts with Viennese quantum community were vital for forming of my later views. I would always remember my conversations with Helmut Rauch during many conferences in V\"axj\"o and my visits to the Atom Institute.  He had his own picture of  the quantum  phenomena, a kind of reality of both the wave function and particle. Helmut Rauch strongly supported the V\"axj\"o series of conferences and he attracted to V\"axj\"o the top experimentalists from Vienna. I spent a lot of time with Johann Summhammer, both in Vienna and during his visits to V\"axj\"o and I was really astonished by his idealist perspective of the physical world. Spending time with him generated first doubts in my ``naive realism''. I still remember that once I found Johann staying near  V\"axj\"o lake and looking at it. He told me something as  ``Look to this beautiful picture which is created in our brain by integrating the clicks of photo-detectors''.

I was lucky to visit the University of Vienna during realization of the final Bell tests and enjoyed the atmosphere of coming great scientific event. I spent a few months with Zeilinger's group during both experiments closing the last loopholes, in  2013 when 
the detection efficiency loophole was closed \cite{Giustina0} and in 2015 when the previous experiment was completed to the loophole free experiment \cite{Giustina}. This is the good place to highlight the contribution of Marissa Giustina in these breakthrough experiments. 
We spent a lot of time by drinking tea and coffee and speaking about the experiments. I was curious in understanding ``experimental technicalities''. (In fact, I had my vested interest, I was searching for experimental data that might show some kind of deviations from quantum theory.) I was really surprised by Marissa's experience in the design and methodology of experimenting in quantum foundations and 
her remarkable understanding of the basic problems of quantum theory. She worked hardly on realization of both experiments and spent a few hundred hours in the basement of Vienna Hofburg Palace (for the last experiment) by collecting  data and then analyzing these huge data-sets and modifying the experimental setup.  
Definitely her contribution into closing all loopholes in the Bell test was crucial. As a university professor, I was also interested in the pedagogical practice of Anton Zeilinger: he put young person in the heart of this fundamental experiment. And Marissa justified his hopes.       

Relaxing conversations with Anton Zeilinger and the members of his experimental group, especially with Sven Ramelow, Rupert Ursin,  Johannes Kofler, and Bernhard Wittmann were vital for my leave-taking from ``naive realism''. In particular, I was excited by Anton's stories regarding his debates with Dalai Lama. I suspect that recording of these debates has never decrypted and published. And I think that it should be done, soon or later. In Vienna I learned that a photon is simply a click of photo-detector and nothing more, that the correlations in Bell experiment are natural and can be influenced by a variety of technicalities, that some parameters of experimental tests cannot be defined on the basis of quantum theory alone but depend on a lot of experience of the experimentalists who select them by hands. (Joint article \cite{KHRR} is nice scientific memory about the days spent with Zeilinger's group.)
Consequently, experimentalists are less orthodox than theoreticians and more open for beyond-quantum theories, especially if such theories can be helpful for designing experiments. After my talk at one of the V\"axj\"o conferences, Alain Aspect asked me a question about negative probability, its possible interpretations and description of the Bell argument. I was surprised, since I didn’t  expect such question from experimentalist. Moreover, by getting to know that I have a series of papers and books on negative probability and its interpretation (e.g., \cite{INT}), he continued the discussion during the lunch and told about his own model with negative probabilities violating the Bell inequality, later he sent me this paper, old and practically unknown, cf. with Zeilinger's 
paper \cite{Zeilinger}.  
\\ 

It is interesting that the intensity of the debates in  V\"axj\"o is going down. The last hot confrontation was between Ozawa and Lahti on Ozawa's rethinking of Heisenberg's uncertainty relation and the possibility of its violation \cite{O4}. The cause of debating decline is not clear. It might be that the basic quantum foundational problems have been resolved or at least the community came to the consensus. But it might be that one part of the quantum community with its specific perspective on foundations became so powerful that this led to disappearance of outsiders from the scene, at least in V\"axj\"o. It might be also a signal of decreasing of interest to quantum foundations. This is the good place to recall that in 1970s the interest to foundations of QM wasn't so high.
During the PhD-defense of Alain Aspect one of jury's members made the remark: but, where would this bright student find a job? So, Bell's test strongly stimulated foundational research. Successful loophole free experiments made the widespread illusion that,  roughly speaking,  there is no any way beyond QM and it is frozen forever. I do not think so. One of the messages of this  note is that before trying to go ``beyond quantum'', one should analyze in more detail a variety of possible ways.
And one should remember that the main impact of Bell's model was the possibility of its experimental verification. So, new quantum foundational revolution would be connected to new experimental tests.  As an example of such tests, I point to Sorkin's
equalities. Weihs' group in Innsbruck spent a few years by trying to find violations of these inequalities \cite{Sorkin1,Sorkin2}, but they confronted with too difficult technical problems for detectors' functioning. Up to now, Sorkin's inequalities have not been violated.
\\
 
Experimentalists working on quantum foundations need new theories and their experimental tests as soon as possible, otherwise people would move to the rapidly growing field of quantum technologies ...

\section{From realism to contextuality} 
\label{2}
 
Realism was so deeply in my mind that, for me, it was difficult to dismiss it and imagine the acausal universe driven by irreducible quantum randomness which was advertized by von Neumann \cite{VN}. I spent a few years of my life by working on the mathematical models combining realism and locality and constrained by the violation of the Bell inequalities. I interacted with practically all leading experts in foundations who work on similar models, e.g., Accardi, Kupczynski, Hess and Philipp, De Raedt and  Michielsen, Nieuwenhuizen, Volovich,
de la Pena and Cetto. In spite of controversy in the evaluation of such studies which was culminated in the famous bet of Accardi with Gill, I have the impression that local realistic treatment of quantum phenomena is possible.  I think that it can be done by discarding hidden assumptions not only of the Bell inequality project, but of quantum theory as whole, e.g.,  {\it the ergodicity assumption}
 \cite{KHRERG}.
 
From my viewpoint, the main problem of the local realistic approach to quantum phenomena is its clustering into special models working for concrete phenomena (as at least claimed by the authors) without general theory. For example, the probability interference can be modeled under the assumption that there exists indivisible quantum of time.
\\

My leave-taking from realism, at least in its naive form, started with studying  Bohr's works  and discussions with Plotnitsky on the complementarity principle. After a while  I understood,  that the complementarity principle is seeded in contexuality of measurements. Contextuality is understood very widely here, as dependence on the complex of experimental conditions. According to Bohr, measurements' outcomes can't be treated as objective properties of a system; they are generated in the complex process of interaction between a system and a  measurement device. Bohr operated with the term ``conditions'', but not contexts. In the modern terminology, these are experimental contexts. Such notion of contextuality is wider than so-called joint measurement contextuality which  was invented by Bell in the discussion on possible interpretations of the violation of his inequality and nowadays it is intensively studied in quantum information theory. We recall that Bell's contextuality is about measurement of an observable $A$ in the contexts of joint measurements with other observables, say $B_1,...,B_n$ which are compatible with $A.$ As well as Bohr,  Bell did not use the term contextuality.  

During the recent years I worked a lot on  Bohr's contextuality and its place in the formulation of the principle of complementarity \cite{NL0B,ABell}. This is the basic principle of QM. It is typically reduced to wave-particle duality. However, careful reading of Bohr shows that this principle has a more complex structure. It might be better to call it
{\it the principle of contextuality-complementarity.} This is an epistemic principle on structuring knowledge which can be obtained about a quantum system. The ontological basis of the complementarity principle is the quantum postulate  declaring the existence of indivisible quantum of action given by the Planck constant $h$ \cite{BR28,BR29,BR29a}. In articles \cite{HP,HP1},  the direct  epistemic counterpart of the quantum postulate is formulated in the form
of {\it the quantum action invariance.} This principle and the complementarity principle are treated as two basic assumptions  of QM, similar to Einstein's principles of special relativity (cf. with Zeilinger's search for fundamental principles of QM \cite{Zeilinger}).  
\\

The Bohr complementarity and Heisenberg  uncertainty principles are deeply interconnected. Heisenberg also related his principle to the quantum postulate. This is the good place to remark that recently the Heisenberg  uncertainty principle  was reconsidered by Ozawa \cite{O4,O5} within theory of quantum instruments.
\\

Now we come to the issue of {\it quantum nonlocality.}  As well as local realism, this is an ambiguous notion. Two different sorts of  nonlocality are often mixed. One of them can be called intrinsic quantum nonlocality. This is apparent nonlocality of quantum state update based on the projection postulate,  another nonlocality is nonlocality of the hidden variables models violating the Bell inequalities. These two sorts of nonlocality should be sharply distinguished, as was done in Aspect's talk   at one of the V\"axj\"o conferences. He started his lecture with the statement that QM is intrinsically nonlocal, since, for a compound system, the operation of the state projection resulting from a local measurement is nonlocal. Then he told that this nonlocality is so unpleasant for any physicist that one should search for another theoretical formalism which would be free of projection nonlocality. And in this way one introduces hidden variables. Finally, he tells that, as is shown by the violations of the Bell inequalities, the subquantum locality (Bell locality) also should be rejected. This is very consistent presentation the Bell inequality project  \cite{AA1}, much better than original Bell's motivation - to explain the perfect quantum correlations \cite{Bell1}. From my viewpoint, correlations are not mystical at all, irrelevantly to their quantumness or classicality (see \cite{ENT,ENT1} on probabilistic entanglement of observables and 
\cite{AKNJ} on classical Brownian entanglement).  I prefer Aspect's motivation for going beyond QM as attempting to find a deeper theory which would not suffer of projection nonlocality.
\\

In paper \cite{NL1a} projection nonlocality is analyzed in connection to selection of interpretations of QM. As is well known, QM is characterized by the diversity of interpretations. Some experts consider it as a sign of the crisis in quantum foundations (e.g., as Ballentine at the first V\"axj\"o conference in year 2001), others rigidly keep to one concrete interpretation without even paying attention to the existence of other interpretations. But the majority of researchers working in quantum physics are not interested in the  interpretation problem.  I have asked the question:{\it What is your interpretation of the wave function?}
to hundreds of participants of the V\"axj\"o conference series. Some of them were not able to say something consistent and others  automatically replied: {\it I use the Copenhagen interpretation!} However, the meaning of the latter is another delicate problem. This is a good place to recall Plotnitsky claiming that the term  ``Copenhagen interpretation'' unifies a bunch of interpretations; he suggested to speak about the interpretations in the spirit of Copenhagen \cite{PL0,PLRWR}. My experience of discussions in  V\"axj\"o supports this suggestion.
\\

Following Ballentine \cite{BL,BL2,BL2a}, we cluster the interpretations into two classes:
\begin{itemize}
\item {\bf SI} the statistical (ensemble) interpretation,
\item {\bf II} the individual interpretation.
\end{itemize}
By  {\bf SI} a quantum state describes the statistical features of an ensemble of identically prepared systems; by  
{\bf II} a quantum state is a state of an individual system, say an electron. Ballentine identifies {\bf II} with the orthodox
Copenhagen interpretation.  Interpretation discussions, {\bf SI} vs. {\bf II}, flared up during a few conferences, Ballentine and
 Nieuwenhuizen advocated consistently {\bf SI} and Lahti gave a few firing speeches in favor of {\bf II}.

The projection nonlocality issue depends on an interpretation. By following {\bf II} one really confronts
(as, e.g., Aspect) with the intrinsic quantum nonlocality. However, by following {\bf SI} one treats the projection postulate as the quantum analog of the Bayes rule for classical probability update. 

In fact, the EPR paper \cite{EPR} was directed against  the orthodox Copenhagen interpretation. Unfortunately, it was not formulated precisely. Later in 1949 Einstein pointed out \cite{EPR49} that there are two choices:
\begin{itemize}
\item a) the orthodox Copenhagen interpretation and  nonlocality, 
\item b) the statistical (ensemble) interpretation and locality. 
\end{itemize}
We note that Einstein considered the intrinsic quantum nonlocality - projection nonlocality. 

By struggling against the orthodox Copenhagen interpretation Einstein did not question the validity of quantum theory. Specially the reasoning in the EPR-paper \cite{EPR} is crucially based on the validity of the Heisenberg uncertainty principle. I claim that Bell's model with hidden variables contradicts this principle as well as the complementarity  principle. To follow Bell, one should from the very beginning reject these principles and also the quantum postulate, the existence of the indivisible quantum of action. This would be the difficult decision, similar to the decision to reconstruct the special relativity by rejecting the Einstein postulate on the constancy of the light velocity.

What is the main problem of the Bell-like  models with hidden variables? This is the identification of the quantities of a subquantum model with 
quantum observables, the  identification of their outcomes and probability distributions. This casualty of the Bell project was highlighted by De Broglie \cite{DeBroglie}, immediately after publication of the Bell works. Unfortunately, his message was completely ignored. De Broglie interpreted his double solution theory as a hidden variables theory. (We remark that so common identification of De Broglie's and Bohm's  theories is wrong. In particular, De Broglie did not consider his theory as physically nonlocal; he just pointed to apparent nonlocality of mathematical equations.) For De Broglie, the values and probability distributions of subquantum quantities and quantum observables 
should not be identified. Hence, in subquantum theories  not only variables, but even quantities (functions of hidden variables 
$\xi= \xi(\lambda))$ and their probability distributions  are hidden. 

De Broglie's approach to subquantum theories matches perfectly with 
the {\it Bild conception}  for structuring of scientific theories which was established by Hertz and Boltzmann and later applied by  Schr\"odinger to QM \cite{Hertz}-\cite{SCHB} (see also \cite{KHRBild} for recent development of the Bild conception). By this conception one should distinguish an observational model  and a causal theoretical model. As was pointed out by Schr\"odinger, QM is an observational model and it is characterized by acausality (as was emphasized by von Neumann \cite{VN}). One can search for causal theoretical models beyond QM.  Due to Hertz and Bolzmann, coupling between causal theoretical and observational models  can be tricky and varying. From this viewpoint, Bell suggested just one special coupling. Hence, employing the Bild conception opens the door to creation of a variety of subquantum causal theoretical models, in spite the negative output of the Bell project. Unfortunately, in quantum foundations the Bild conception is practically unknown. Its partial resemblance can be found in ontic-epistemic structuring of physical theories due to Primas and Atmanspacher \cite{ATM0,ATM}. 

A promising causal theoretical model for QM is {\it prequantum classical statistical field theory}, see \cite{Beyond}. This is model of classical random fields reproducing quantum averages and correlations. Its coupling with QM is more tricky than for the Bell model with hidden variables. 

Bohr's contextuality is formalized within the contextual measurement model (CMM) developed in \cite{KHROPEN}.  CMM is based on the calculus of contextual probabilities which closely related to generalized probability models. CMM serves as the basis for the {\it V\"axj\"o interpretation} \cite{V2} of QM, a contextual statistical interpretation. By this interpretation QM is the machinery for  generally non-Bayesian probability update. One of the fundamental exhibitions of non-Bayesianity is the violation of the classical {\it formula of total probability}  (FTP) \cite{KHR_CONT}. Contextual generalization of FTP contains an additional term, the interference term. In QM this is the formula for interference of probabilities, e.g., in the two slit experiment (cf. Feynman \cite{FeynmanP,Feynman}).  
Paper \cite{NL} exposes purely quantum approach to the Bell inequalities which are exemplified as the tests of incompatibility of local variables, or in the mathematical terms, as the tests of noncommutativity: $[\hat A_1, \hat A_2]\not=0$ and $[\hat B_1, \hat B_2]\not=0$ for Alice's and Bob's observables, respectively.

Recent article \cite{QMasanao} is devoted to critical analysis of QBism's private experience perspective in light of Ozawa's intersubjectivity theorem \cite{OIT} based on theory of quantum instruments and the indirect measurement scheme. This paper is continuation of my challenging QBism during the last 20 years \cite{KHRQBism1,KHRQBism2}. My intersubjectivity paper was carefully analyzed and responded by the QBism community \cite{QBismReply1,QBismReply2}. But neither I nor Ozawa are satisfied by QBists's replies and the discussion will continue, may be at V\"axj\"o 2024 conference.   

I was one of the pioneers in applications of the methodology and formalism of quantum theory outside of physics \cite{KHC2,KHC3,QL0,QL1}. This area is known as {\it quantum-like modeling}. I put a lot of efforts to its establishing and advertising and I am thankful to all my friends contributed to the volume on {\it quantum-like revolution}  \cite{Zeit}. It is important to note that, in spite of my opposition to QBism in quantum physics, I appeal to it \cite{QB_DM} as a possible interpretational basis for modeling cognition and decision making with quantum formalism. I can't say that the idea of applying the quantum methods outside of physics was immediately welcomed by the participants of the first V\"axj\"o conferences. I remember some hating comments on my talks devoted to quantum-like modeling  at the beginning of this century: ``Please don't touch our beautiful quantum theory  with your dirty hands!'' But by opening the Overton  window wider and wider I was able to smooth the reactions of ``rigid quantum physicists'' on the use the quantum methods outside of physics, at least in decision making. Later a few special sessions on quantum-like modeling in decision making, economics and  finance were organized in cooperation with Emmanuel Haven who contributed so much in advertising this area of research in V\"axj\"o. Nowadays quantum-like modeling is the well established area of research. One of its problems is attraction of numerous newcomers from various domains of science  who are not so well educated in quantum theory. Often  
 ``Aladdin and the Magic Lamp''. But this is merely the illness of too rapid growth.         

I was lucky to speak as well with Kolmogorov, concerning foundations of probability and possible generalizations of his axiomatics,  as with Mackey regarding the conditional probability structure of QM. Then this viewpoint on quantum probability was discussed with Accardi, Ballentine, and Gudder and these discussions were imprinted in my mind. My views on the causes of the Bell inequalities violations were influenced by conversations with Lochak, De Baere, and De Muynck.
\\

My works in quantum foundations can be interesting for all who are not excited by quantum mysteries and paradoxes (or scandals 
\cite{TheoQLR}) and consider QM as an observational model describing quantum measurements. QM without nonlocality and spooky action at a distance need not seem insipid. One can find a plenty of interesting problems: theoretical, experimental, mathematical, and philosophical. QM treated as a measurement model, works very well and there is no reason to question it. Its apparent acausality is typical for observational models as was highlighted already by Hertz, Boltzmann, and  Sch\"odinger \cite{Hertz}-\cite{SCHB}. If one wants to go beyond QM, it is natural to appeal to the Bild conception. In particular, the violation of the Bell inequalities should be analyzed within this conception.

\end{document}